# TOWARDS AXIOMATIC FOUNDATIONS FOR CONCEPTUAL MODELING: AN EXAMPLE

*arXiv - Preprint*

Peter Fettke, Saarland University, Saarbrücken, Germany, and German Research Center for Artificial Intelligence (DFKI), Saarbrücken, Germany, email: peter.fettke@dfki.de

Wolfgang Reisig, Humboldt-Universität zu Berlin, Berlin, Germany, email: reisig@informatik.hu-berlin.de

## Abstract

*Conceptual modeling is a strongly interdisciplinary field of research. Although numerous proposals for axiomatic foundations of the main ideas of the field exist, there is still a lack of understanding main concepts such as system, process, event, data, and many more. Against the background of the tremendously gaining importance of digital phenomena, we argue that axiomatic foundations are needed for our discipline. Besides the general call, we provide a particular case study using* HERAKLIT. *This modeling infrastructure encompasses the architecture, statics, and dynamics of computer-integrated systems. The case study illustrates how axiomatically well-founded conceptual models may look like in practice. We argue that axiomatic foundations do not only have positive effects for theoretical research, but also for empirical research, because, for instance, assumed axioms can explicitly be tested. It is now time to spark the discussion on axiomatic foundations of our field.*

*Keywords: systems composition, data modelling, behavior modelling, composition calculus, algebraic specification.*

> *"Formalization is important to economics, because it allows readers and others to identify the precise assumptions that underpin any purported conclusion, to verify that the assumptions really do imply this conclusion, and to check how deviations from the assumptions might alter the conclusion."*
> Paul Milgrom (2017), p. 3f., Nobel Laureate 2020

# 1 The Field and Branches of Conceptual Modeling

As numerous other terms in computing disciplines, the term "conceptual modeling" is used to denote a particular kind of *problem*, a possible *solution* to that problem, and the *academic field* studying that problem and its known solutions:

(a) Conceptual modeling as a *problem*: Concepts can be understood as intellectual building blocks. They are important for thinking, reasoning, and describing as well as for designing, manipulating, or acting in the world. Concepts are not only used by humans but are also needed for communication with machines, e.g. for describing the functionality a running program offers as well as for denoting steps, procedures, and variables of an algorithm formulated in a particular programing language. Although no consensus on the nature of concepts exists, in general, it can be said that conceptual modeling aims at providing a *shared understanding* of a particular domain of interest (cp. Uschold, p. 96).

(b) Conceptual modeling as a *solution*: Conceptual modeling is used in many different contexts, e.g. database modeling, software development, business process reengineering, workflow modeling, software selection, configuration of enterprise resource planning systems, and many more. Both in academia and industry, numerous approaches to conceptual modeling emerged. Some speak of hundreds or even thousands of different approaches. The *International Standardization Organization* (ISO), *Object Management Group* (OMG), or other institutions standardized several approaches, e.g. *Business Process Modeling Notation* (BPMN), *Unified Modeling Language* (UML), or *Petri nets*. Others are de facto standards since market-leading tools are available supporting particular solutions, e.g. *Architecture for Integrated Information Systems* (ARIS) by Software AG and SAP, or *Directly-Follow-Graphs* used by leading process mining vendors. In general, each approach claims to provide a solution for a particular problem of conceptual modeling.

(c) Conceptual modeling as an *academic field* of inquiry: Having in mind the wide and heterogeneous spectrum of different problems and solutions of conceptual modeling, it is not surprising that the field is interdisciplinary by nature. Although conceptual modeling is viewed as a niche topic in the information systems discipline by Wand and Weber (2017) and others, it is undisputable that conceptual modeling has longstanding research traditions outside this discipline, too (Härderer and Fill (2021), Hirschheim et al. (1995)). Hence, we conclude that conceptual modeling is a field of its own, consisting of several research branches rooted in different academic disciplines. Some important branches of the field and their major outlets for publishing results are: business process management (*International Conference on Business Process Management*, since 2003), distributed systems and Petri nets (*International Conference on Application and Theory of Petri Nets*, since 1980), database systems (e.g. *International Conference on Conceptual Modeling*, since 1979), business informatics and enterprise modeling (*Enterprise Modeling and Information Systems Architectures*, since 2005, *The Practice of Enterprise Modeling*, since 2008), and software engineering (e.g. *International Conference on Model Driven Engineering Languages and Systems*, since 1998). Additionally, several areas of theoretical computer science, e.g. automata theory, and formal languages, are of central importance for the field as a whole, too.

The problems, solutions, and branches of conceptual modeling give a rich picture of different academic ideas important for understanding and designing the world we live in. Against the background of recent developments, Recker et al. (2021) started the discussion on important assumptions of the field from the perspective of the information systems discipline. We argue that this important discussion should be extended in at least two directions: Firstly, it is open whether the identified assumptions are generally accepted in the whole field of conceptual modeling or are only accepted in particular branches of conceptual modeling. Secondly, we like to deepen the ongoing discussion and direct it to important core phenomena of conceptual modeling (cp. Alt (2013), Weber (1997)). In particular, we pose the question: Are there more basic assumptions about core phenomena conceptual modeling is dealing with? How might the understanding of these core phenomena might change in the digital world? For example, central terms of our field such as *concept*, *data*, *process*, *object*, (*sub*-) *system*, and many more are not uniformly defined and understood. In addition, the phenomena related to these terms are not neatly separated from one another but are highly interwoven, both in academia and practice, e.g. there is no sharp boundary between the "data" and the "process" in a "system". Data, processes, and systems in the world are highly interrelated and integrated, even with a higher degree in the age of "digital first".

In this position paper, we argue that besides bold and far-reaching speculations, also precise and explicit foundations are needed for the progress of our discipline. Conceptual modeling as a scientific field should not only speculate on important ideas but should also precisely put forward assumptions and derive interesting conclusions which can be verified by others. Particularly, for this end, we focus on the central idea of the *axiomatic method* for foundations for conceptual modeling. In doing that, the next Section summarizes the main idea of the axiomatic method, Section 3 sketches an example in our domain. The paper closes with some conclusions in Section 4.

## 2      The Axiomatic Method in Empirical Sciences

The idea of axiomatizing a body of knowledge is anything but new and can be traced back, at least, more than 2,000 years to *Euclid* who gives axiomatic foundations for geometry (Suppes (2002)). Since then, tremendous progress has been achieved, e.g. the flourishing field of geometry develops in different branches with many important applications and different foundations, e.g. (non-) Euclidian geometry.

Besides particular applications in geometry, the idea of axiomatizing develops in different directions and is differently understood today. One widely accepted and well-practiced approach can be understood as an *informal axiomatizing*. This approach is attributed as *informal* because it does not use a formal system to formulate the main ideas of a body of knowledge. The framework used is based on *intuitive set-theory* and combines formal with informal ideas (Suppes (1957)). As such, it is always claimed that the formulated ideas can be expressed in the formal system of *first order predicate logic*. However, in everyday scientific practice, just informal language is used.

The idea for this kind of axiomatizing can be traced back to *model theory*, as developed in logic and mathematics. Particularly, the *Bourbaki* group applies this idea to overcome the basic crisis in mathematics at the beginning of the last century (Bourbaki (1957)). However, in the meantime, this approach is the standard approach not only used in mathematics, but also in empirical sciences which numerous examples clearly and impressively demonstrate. For instance, Suppes (2002) provides examples from physics, philosophy, psychology, computer science, economics, and semiotics.

The core idea of an informal axiomatizing is to introduce a *mathematical structure S* consisting of:

- *Basic sets*: The basic sets define the basic objects of the structure *S*, e.g the symbol *M* denotes a basic set.
- *Derived sets*: Several relations can be defined on basic sorts or derived sets, e.g. the symbol < denotes a binary relation on *M*.
- *Functions*: Mathematical functions can be defined as operations on sets, e.g. the symbol *f* denotes a function *M* on itself.
- *Properties*: Certain properties hold for all elements of the structure *S* ("axioms"), e.g. for all *a, b* element of *M* holds: *a < b* or *b < a*.

The axiomatic method is already intensively used in conceptual modeling. Most, if not all *computer science*-oriented branches of conceptual modeling apply the axiomatic method. But, in *management*-oriented approaches, this idea is already used, too, which is not only demonstrated by the quote of the Nobel Laureate Paul Milgrom cited in Section 1. One of the most prominent examples stemming from the information systems discipline is the framework introduced by Wand and Weber (1990) which is based on Bunge's work. Although many contributions relying on this framework do not explicitly discuss or contribute to the axiomatic foundations of conceptual modeling, it is indisputable that Wand and Weber (1990) apply the axiomatic method. To sum up, up to now, the axiomatic method has a longstanding academic tradition and it is fully justified that it is not only of major importance for science in general but also for the field of conceptual modeling in particular.

*Towards Axiomatic Foundations for Conceptual Modeling*

# 3 Example

Next, we illuminate the axiomatic method by an example that sketches some core ideas of HERAKLIT. HERAKLIT is an infrastructure for conceptual modeling which particularly addresses the requirements for modeling computer-integrated systems in the age of "digital first". A HERAKLIT model of the world consists of three pillars:

- *Architecture*: Today's information systems are composed of several sub-systems. A *module* is the basic building block for composing complex systems.
- *Statics*: Structures describe symbolic *objects* as well as objects in the real or imaginary world, e.g. customers, receipts, products, invoices, organizations, employees etc.
- *Dynamics*: Changes in the world are described by a flow of *events*.

All three pillars are strongly integrated as is demonstrated by the following example describing a restaurant business with several branches. All branches are structured according to the same scheme and behave according to the same patterns; however, they differ in some details. The scheme of all branches, an example of a branch, and a run in this branch are modeled.

A branch is composed of three modules, namely *entry*, *dining area*, and a *kitchen* (Figure 1a). The major objects such as *clients, tables, orders,* etc. are defined by a signature-structure (Figure 1b). A particular instantiation for a branch depicts Figure 1c. Figure 1d shows a particular flow of events in the restaurant: *Alice* and *Bob* enter the restaurant, order a meal, the meal is concurrently prepared and cooked, and served to both customers. After having the meal, the customers leave the restaurant.

Note, that all concepts of HERAKLIT and all concepts of this example are introduced axiomatically. More technically, HERAKLIT integrates well-known concepts, namely, the idea of model theory and axiomatic specification (statics), Petri nets (dynamics), and composition calculus (architecture). More technical details and case studies are described by Fettke and Reisig (2021a, b).

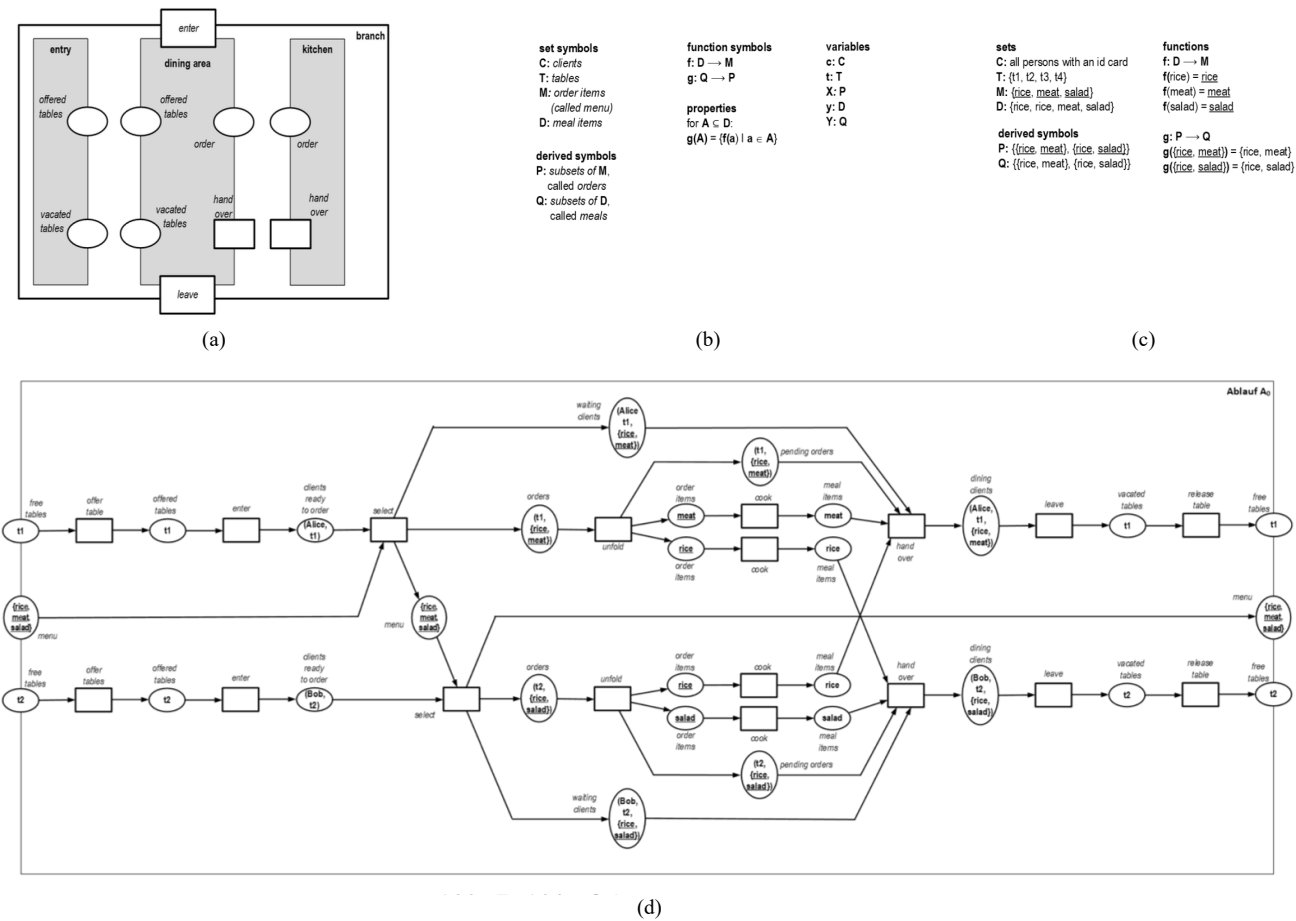

*Figure 1.  Axiomatic modeling: an example based on HERAKLIT.*

# 4 Conclusions

This position paper emphasizes the intellectual power of the axiomatic method in general and for the field of conceptual modeling in particular. We strongly agree on the positive aspects of formalization mentioned in the quote from Nobel Laureate Paul Milgrom (cp. Section 1). Furthermore, we fully share the surprise and wonder described by Weber (1997) on using formalization in our field:

"[W]e [Wand & Weber] have often been criticized for the formal approach we have used to articulate our models. I am perplexed by these criticisms because I cannot conceive of a discipline that is serious about its foundations if it proscribes use of mathematics to articulate these foundations." (p. 33)

Looking back the last decades, it is safe to say that many axiomatic foundations introduced by Wand and Weber are well-accepted as well as often empirically tested, particularly in the information systems branch of conceptual modeling. However, other branches rely on different axiomatic systems. Hence, we believe that it is now time to intensify the discussion on the axiomatic foundations of our field.

Particularly against the background of digitization, this is of major importance. Most, and foremost, it is necessary to explicate what is meant by digital phenomena, e.g. digital data, digital computers, and digital ecosystems. Since several branches of conceptual modeling already started the discussion on how conceptual modeling can embrace digital transformation, now a deeper discussion is needed.

Besides the major advantages of the axiomatic method, we do not argue for the abandonment of other forms of scientific inquiry. The axiomatic method can easily be integrated with other methods such as bold speculations and empirical approaches like experiments and case studies. With this in mind, we strongly believe in a prosperous future of our field.

# References

Alt, S. (2017): Work System Theory: Overview of Core Concepts, Extensions, and Challenges for the Future. *Journal of the Association for Information*, 14(2), 72-121.

Bourbaki., N. (1950): The Architecture of Mathematics. *The American Mathematical Monthly*, 57(4): 221-232.

Fettke, P., Reisig, W. (2021a): *Modelling service-oriented systems and cloud services with HERAKLIT*. Advances in Service-Oriented and Cloud Computing. Communications in Computer and Information Science (CCIS), Springer, 77-89.

Fettke, P.; Reisig, W. (2021b): *Handbook of HERAKLIT*. HERAKLIT working paper, v1.1, September 10, 2021, http://www.heraklit.org

Härer, F., Hans-Georg Fill, H.-G. (2020): *Past Trends and Future Prospects in Conceptual Modeling - A Bibliometric Analysis.* 39th International Conference on Conceptual Modeling 2020: 34-47.

Hirschheim, R., Klein, H. K., & Lyytinen, K. (1995). *Information Systems Development and Data Modeling: Conceptual and Philosophical Foundations*. Cambridge University Press.

Milgrom, P. (2017): *Discovering Prices - Auction Design in Markets with Complex Constraints*. New York: Columbia University Press.

Recker, J., Lukyanenko, R., Jabbari, M., Samuel, B. M., & Castellanos, A. (2021). From Representation to Mediation: A New Agenda for Conceptual Modeling Research in a Digital World. *MIS Quarterly*, *45*(1), 269-300. https://doi.org/10.25300/MISQ/2020/16207

Suppes, P. (2002): *Representation and Invariance of Scientific Structures*. Chicago: CSLI Publications.

Suppes. P. (1957): *Introduction to Logic.* New York et al: Van Nostrund Reinhold.

Uschold, M., Gruninger, M. (1996): Ontologies: principles, methods and applications. *The Knowledge. Engineering Review,* 11(2), 93-136.

Wand, Y., Weber, R. (2017): Thirty Years Later: Some Reflections on Ontological Analysis in Conceptual Modeling. *Journal of Database Management*, 28(1), 1-17.

Wand, Y., Weber, R. (1990): An Ontological Model of an Information System. *IEEE Transaction on Software Engineering*, 16(11): 1282-1292.

Weber, R. (1997): *Ontological Foundations of Information Systems*. Melbourne: Coopers & Lybrand.